\documentclass[prl,aps,twocolumn,showpacs]{revtex4}

\usepackage{epsfig}
\pdfoutput=1
\usepackage{hyperref}
\begin{document}


\title{The History of Inflation from Microwave Background Polarimetry and Laser Interferometry}


\author{Jerod Caligiuri}
\author{Arthur Kosowsky}
\affiliation{Department of Physics and Astronomy, University of Pittsburgh, Pittsburgh, PA 15260 USA}
\affiliation{Pittsburgh Particle Physics, Astrophysics, and Cosmology Center (PITT PACC), Pittsburgh, PA 15260 USA}
 
\author{William H.~Kinney}
\affiliation{Department of Physics, University at Buffalo, Buffalo, NY 14260 USA}

\author{Naoki Seto}
\affiliation{Department of Physics and Astronomy, Kyoto University, Kyoto 606-8502 JAPAN}


\date{March 4, 2015}

\begin{abstract}
A period of inflation in the early Universe produces a nearly scale-invariant spectrum of
gravitational waves over a huge range in wavelength. If the amplitude of this gravitational wave
background is large enough to be detectable with microwave background polarization measurements,
it will also be detectable directly with a space-based laser interferometer. Using a Monte Carlo sampling of 
inflation models, we demonstrate that the combination of these two measurements will strongly 
constrain the expansion history during inflation and the physical mechanism driving it. 
\end{abstract}


\maketitle


\section{I. Introduction}
The pursuit of comsological B-mode microwave background polarization \cite{Kamionkowski:1996ks,Zaldarriaga:1996xe} at large angular scales has seen a burst of attention in the past year since the BICEP2 experiment  \cite{Ade:2014xna} announced the possibility of a detection in March 2014.  Further investigation has revealed that the dominant source of this signal is actually galactic dust \cite{Flauger:2014qra,Mortonson:2014bja, PlanckDust2014, BICEP_Planck2015} rather than being cosmological in nature.  This does not rule out the existance of cosmological B-modes, but it is now clear that detecting a cosmological B-mode signal at significantly lower amplitude than the BICEP2 signal requires better characterization of foreground polarization signals via multi-frequency measurements. If we ultimately measure a primordial B-mode signal, it may provide an otherwise unobtainable window into physics at extremely high energies and the evolution of the Universe during its earliest moments.

A period of exponential expansion in the early Universe, known as inflation, resolves several observational difficulties with the standard cosmological model (e.g., \cite{Guth:1980zm,Linde:1981mu,Albrecht:1982wi,Kazanas:1980tx,Starobinsky:1980te,Sato:1981ds,Sato:1980yn,Mukhanov:1981xt,Mukhanov:2003xw,Linde:1983gd}). Such a period of expansion can be driven by some component with a nearly constant energy density, such as the potential energy of an effective scalar field (the ``inflaton"). During this epoch, quantum fluctuations in the inflaton field are amplified into classical density perturbations which grow by gravitational instability into the structures visible in the Universe today. Likewise, quantum fluctuations in the tensor components of the inflating spacetime
are amplified into a stochastic background of gravitational waves, with a nearly scale-invariant power spectrum and an amplitude depending on the energy scale of inflation. 
Once generated, these tensor perturbations propagate almost freely through the Universe, their wavelengths increased by the expansion of the Universe and their amplitudes decreased by the same factor once the wavelength comes inside the horizon. 

The hallmark of an inflationary gravitational wave background is its extremely wide range in wavelengths, roughly from the scale of the horizon today down to terrestrial scales. 
If inflation occurred when the temperature of the Universe was around the Grand Unification scale of $10^{16}$ GeV, the gravitational wave background will produce a pattern of B-mode polarization large enough to be seen by upcoming experiments. Remarkably, an inflationary tensor perturbation signal of roughly this amplitude will likely also be detectable directly with future space-based laser interferometers 
\cite{Turner:1996ck,Ungarelli:2005qb,Cooray:2005xr,Smith:2005mm,Smith:2006xf,Chongchitnan:2006pe,Friedman:2006zt,Kuroyanagi:2014qaa,Jinno:2014qka,Boyle:2014kba}. Direct detection will probe wavelengths that are $15$ orders of magnitude smaller than those inducing microwave background signals;  measurements
at these two scales span a large portion of the observable inflation epoch, and may therefore reveal the physical process driving inflation.Gravitational wave backgrounds of smaller amplitude are more difficult to detect at both cosmological and local scales.  

A recent paper showed that an interferometric detection of the inflationary tensor signal will provide qualitatively new information about inflation, namely a precise measurement of any departure from a pure power-law spectrum \cite{Caligiuri:2014sla}. If we can obtain information about
inflation at widely varying scales, a natural question arises: ultimately, how well can the history of inflation, and the physics driving it, be determined? Here, we determineconstraints on the expansion history during inflation, as well as the corresponding effective potential governing the evolution of the inflaton, from future measurements at both microwave background and interferometer scales. We use a Monte Carlo method to generate random inflation models with an initially slowly varying inflaton value (the so-called ``slow roll" condition) at the time when perturbations on the scale of the horizon today are generated \cite{Kinney:2002qn,Easther:2002rw} and which satisfy current measurements \cite{BICEP_Planck2015}. Of these models, only a very small fraction satisfy tensor amplitude constraints at the few percent level on both scales, and the resulting expansion histories during inflation are strongly constrained. 

Other recent works use similar techniques to constrain inflation. Ref.~\cite{Contaldi:2014rna} considers only microwave background measurements; our results show that the additional information from an interferometric detection greatly restricts the range of acceptable inflation models. Ref.~\cite{Chongchitnan:2006pe} focused on detectability of tensor perturbations in specific models by interferometers of a given sensitivity.
We extend this work to consider the generic features of inflation models satisfying a particular pair of B-mode polarization and interferometer measurements. Ref.~\cite{Smith:2006xf} considered models with the largest interferometer signals. We also consider interferometer measurements with sufficient sensitivity to constrain both the amplitude and spectral index of tensor perturbations at local scales, providing additional ability to discriminate between inflationary expansion histories.

\section{II. Methods}
We compute the dynamical history and resulting scalar and tensor perturbation spectra for random inflation models chosen via the technique of Monte Carlo inflation flow potential reconstruction \cite{Kinney:2002qn,Easther:2002rw}. The dynamics of an inflation model can be written in terms of the value of an effective scalar field
$\phi$ (the ``inflaton") and its potential energy $V(\phi)$; inflation occurs when the energy density is dominated by the inflaton potential energy. The field $\phi$ will
evolve towards the minimum of the potential as inflation progresses. It is convenient to use the value of the effective
inflaton field $\phi$ during inflation as a time parameter; this can be done as long as $\phi$ evolves monotonically to smaller values.
The Hubble parameter during inflation, $H\equiv (1/a) da/dt$ with $a(t)$ the scale factor, can be used to construct a hierarchy of ``slow-roll parameters" \cite{Lidsey:1995np}
\begin{eqnarray}
\epsilon &\equiv& {m_{\rm Pl}^2 \over 4 \pi} \left(\frac{H'(\phi)}{H(\phi)}\right)^2,\cr
\sigma &\equiv& \frac{m_{\rm Pl}}{\pi} \left[\frac{1}{2} \left(\frac{H''}{H}\right) - \left(\frac{H'}{H}\right)^2\right],\cr
{}^\ell\lambda_{\rm H} &\equiv& \left(\frac{m_{\rm Pl}^2}{4\pi}\right)^\ell
\frac{\left(H'\right)^{\ell-1}}{H^\ell} \frac{d^{(\ell+1)} H}{d\phi^{(\ell +1)}}, \qquad \ell \geq 2.
\label{parameters_definition}
\end{eqnarray}
where primes are derivatives with respect to $\phi$ and the Planck mass $m_{\rm Pl}$ sets the energy units. 
The evolution of the parameters during inflation is
determined by a system of first-order linear equations \cite{Hoffman:2000ue,Kinney:2002qn},
\begin{eqnarray}
\frac{d \epsilon}{dN} &=& \epsilon \left(\sigma + 2\epsilon\right),\cr 
\frac{d \sigma}{dN} &=& - 5 \epsilon \sigma - 12 \epsilon^2 + 2 \left({}^2\lambda_{\rm H}\right),\cr 
\frac{d \left({}^\ell\lambda_{\rm H}\right)}{dN} &=& 
\left[\frac{\ell - 1}{2} \sigma + \left(\ell - 2\right) \epsilon\right]
\left({}^\ell\lambda_{\rm H}\right) + {}^{\ell+1}\lambda_{\rm H}.
\label{eqfullflowequations}
\end{eqnarray}
In these equations the time variable is the number of expansion e-folds before the
end of inflation $N\equiv  \ln (a_{\rm end}/a)$ where
\begin{equation}
\frac{d}{dN} = \frac{m_{\rm Pl}}{2\pi^{1/2}} \sqrt{\epsilon}\frac{d}{d\phi}.
\label{N_deriv}
\end{equation}
Given a solution for $\epsilon(\phi)$, $H(\phi)$ can be obtained from the definition of $\epsilon$, which then requires normalization given the observed amplitude of the scalar power spectrum \cite{Ade:2013zuv}.
The effective potential $V(\phi)$ then follows from the equation of motion for $H(\phi)$,
\begin{equation}
H^2(\phi) \left[1-\frac{1}{3}\epsilon(\phi)\right] = \frac{8\pi}{3m_{\rm Pl}^2} V(\phi).
\label{V_eq}
\end{equation}

Particular models of inflation are obtained by choosing initial values for the slow-roll parameters; their dynamical
evolution corresponds to some trajectory in the slow-roll parameter space. In practice,
we truncate the hierarchy Eqs.~(\ref{eqfullflowequations}) above $\ell=8$ (corresponding to a restricted 
subset of exact inflation models) and then evolve the evolution equations until inflation ends, which we take
as the condition $\epsilon = 1$. Once the slow-roll parameters are all determined throughout inflation as a function
of $N$, the primordial comoving curvature perturbation $\delta\rho/\rho$ arising from the scalar perturbations 
and the tensor/scalar ratio $r$ are, to second-order in the slow roll
parameters, given by
\cite{Lidsey:1995np,Stewart:1993bc},
\begin{eqnarray}
\frac{\delta\rho}{\rho} &\simeq& \frac{H}{2\pi m_{\rm Pl}^{\phantom 2}\sqrt{\epsilon}},\nonumber\\
r &\simeq& 16 \epsilon \left[1 - C \left(\sigma + 2 \epsilon\right)\right]
 \label{eqrsecondorder}
\end{eqnarray}
with $C \equiv 4 (\ln{2} +
\gamma) - 5 = 0.0814514$, where $\gamma \simeq 0.577$ is Euler's
constant, and using the WMAP normalization convention for $r$ \cite{Spergel:2003cb}.
Then the perturbation amplitudes as a function of $N$
can be converted to amplitudes as a function of comoving wavenumber $k$ at the end of inflation by
the relation $a(N)H(N) = k$.
Generating scalar and tensor perturbations for a random sampling of inflation models is thus
reduced to choosing random initial points in the slow-roll parameter space. 
The resulting primordial scalar perturbation power spectrum $P(k)\simeq A_S (k/k_0)^{n_{\rm S}-1}$ and tensor power
spectrum $P_T(k)\simeq A_T (k/k_0)^{n_{\rm T}}$ are both approximate power laws on cosmological scales
around $k_0 = 0.05$ Mpc${}^{-1}$, with $r \equiv A_T/A_S$.

We initially generate $2\times10^{4}$ models within an ellipsoidal parameter space region with principal axes $0.952 < n_{\rm S} < 0.981$ and $0.0 < r < 0.1$;
these ranges are consistent with the 95\% confidence region from current Planck \cite{Ade:2013zuv}, WMAP polarization \cite{Bennett:2012zja}, and BICEP2/Keck+Planck microwave background measurements \cite{BICEP_Planck2015}, as well as baryon acoustic oscillation measurements from the Sloan Digital Sky Survey
Data Release 9 \cite{Ahn:2012fh}, the 6dF Galaxy Survey \cite{Jones:2009yz}, and the
WiggleZ Dark Energy Survey \cite{Blake:2011en} (shown in Fig.~5 of Ref.~\cite{Freese:2014nla}), while the running of the scalar spectral index was restricted to be arbitrarily small  ($-0.001 < dn_{\rm S} / d\ln{k} < 0.001$). Note that the most recent Planck release \cite{Ade:2015xua} provides a constraint of $-0.015 < dn_{\rm S} / d\ln{k} < 0.011$ at $95\%$ confidence. We have not adjusted for this allowed range of values as it will not alter the conclusions of this work and by the time the measurements discussed here are achieved, constraints on $dn_{\rm S} / d\ln{k}$ will likely be at least as strong as we are considering. For simplicity, parameter degeneracies have not been considered. More specifically, this is done by initially setting $\phi_0=0$, $H(\phi_0)=1$, and randomly selecting $\epsilon \in [0,0.1]$,  $\sigma \in [-0.1,0.]$, ${}^3\lambda \in [-0.05,0.05]$, ${}^4\lambda \in [-0.005,0.005]$,  ${}^5\lambda \in [-5\times10^{-4},5\times10^{-4}]$, ${}^6\lambda \in [-5\times10^{-5},5\times10^{-5}]$, ${}^7\lambda \in [-5\times10^{-6},5\times10^{-6}]$, and ${}^8\lambda \in [-5\times10^{-7},5\times10^{-7}]$ with a uniform probability distribution. The models are then freely evolved to the end of inflation when $\epsilon=1$. If inflation lasts for the desired $60$ efoldings of expansion, the expansion rate is then renormalized to the amplitude of the scalar power spectrum.  Those models that produce consistent values of $r, n_{\rm S}$, and $dn_{\rm S} / d\ln{k}$ at $N=60$ as stated above are then saved. The full tensor spectrum is then computed by exact numerical evaluation of the mode equation for the cosmological background defined by the selected solution to the flow equations, in place of the slow-roll approximation in Eq.~(\ref{eqrsecondorder}). The corresponding inflation potential $V(\phi)$ is also computed numerically for each model. Models are not restricted to remain slowly-rolling throughout this evolution (as is appearent in Fig.~\ref{FIG1} and Fig.~\ref{FIG2}, where models evolve more rapidly and steepen). In particular, toward the end of inflation (low values of N), a departure from slow-roll is required for inflation to come to an end.

We then select the subset of models consistent with tensor amplitudes at cosmological scales corresponding to a fiducial tensor-scalar ratio of $r = 0.05 \pm 0.001$, a precision obtainable by currently anticipated experiments.

Given the tensor perturbation spectrum from a particular inflation model, we obtain the tensor spectrum in the present Universe using well-known techniques for computing the transfer function for the amplitudes and wavelengths \cite{Smith:2005mm,Smith:2008pf,Watanabe:2006qe,Weinberg:2003ur}, 
assuming the standard $\Lambda$CDM cosmology \cite{Ade:2013zuv}. For simplicity, we assume that the reheating phase after inflation, in which the energy density stored in the kinetic energy of the inflaton field is converted
to a thermal bath of relativistic particles, occurs rapidly on a time scale short compared to the Hubble time. We then compute a signal for both B-mode polarization and laser interferometry from the tensor perturbations in each model. 

The most recent joint analysis by the Planck and BICEP2/Keck collaborations limit the tensor-scalar ratio to $r<0.12$ at 95\% C.L. \cite{BICEP_Planck2015}.
For interferometers, a stochastic gravitational wave power spectrum is often expressed as $\Omega_{\rm GW}(f)$, the fraction of critical density in gravitational waves per unit logarithmic frequency interval. We assume a
particular fiducial model of the local tensor perturbations with an amplitude of $\Omega_{\rm GW} = 8.2\times 10^{-17}$ at a frequency of $f=0.25$ Hz. We then consider a measurement of this amplitude with a 2$\sigma$ standard error of around $8\%$: this corresponds to the
reference strain sensitivity $u\sqrt{S_{\rm base}(f)}$ of the DECIGO effective interferometer design given in Fig.~2 of \cite{Kawamura:2006up}, with $u=1/5$, $3$ years of observation, and a low-frequency cutoff of $0.2$ Hz to minimize the contaminating astrophysical foreground signal from white dwarf binaries. 
We also assume perfect removal of neutron star and black hole binary signals. (One interferometer design attaining these specifications comprises four sets of three detectors with optimal sensitivity around $1$ Hz. 
Each group can effectively generate two independent interferometers by taking appropriate combinations of data streams from the set. Pairs of these groups will have overlapping orbits to facilitate correlation analysis.  Such an experiment is clearly ambitious, but achievable with known technology at a cost comparable to current large physics and astronomy efforts.) A particular inflation model is considered consistent with the fiducial interferometer signal if its amplitude is within the 2$\sigma$ range for $\Omega_{\rm GW}$ over the frequency band from 0.2 to 20 Hz.

\section{III. Results}

\begin{figure*}[t]
\includegraphics[width=6in]{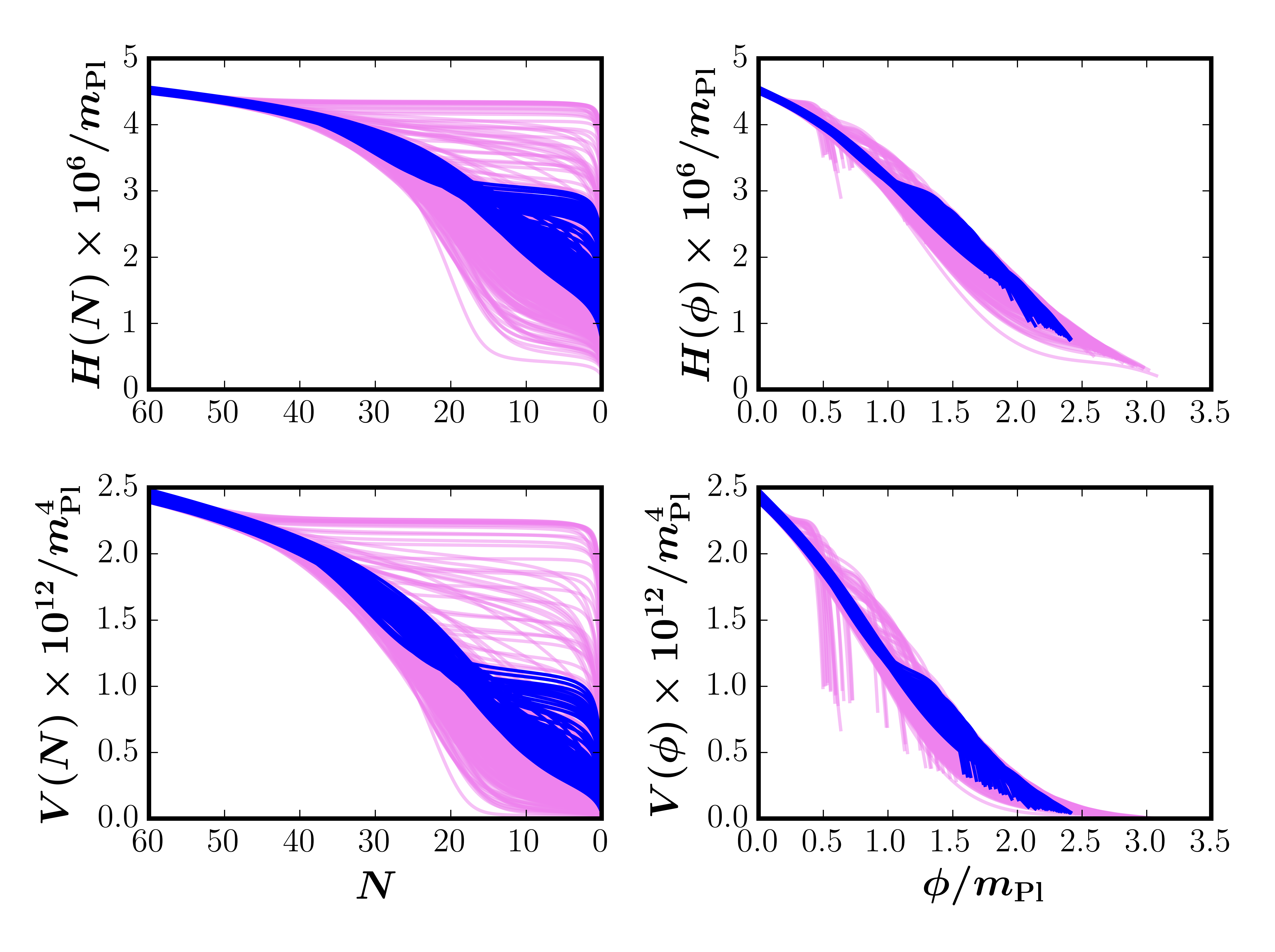}
\caption{The expansion rates in the form of the Hubble parameter, $H$, and the effective potentials driving the evolution of the expansion, $V$, are shown.  All models consistent with $r = 0.05 \pm 0.001$ at cosmological scales are plotted in violet.  Those models that are also consistent at local scales with $\Omega_{\rm GW}(k_0 = 1.6\times10^{14} \rm Mpc^{-1}) = (8.2 \pm 0.69) \times10^{-17}$ are plotted in blue.  Note that modes with wave number $k = 1.6\times10^{14} \rm Mpc^{-1}$ correspond to frequencies $f = 0.25$ Hz. \emph{Right Column}: $H$ and $V$ as a function of the number of efoldings before the end of inflation in the top and bottom panels respectively.  \emph{Right Column}: $H$ and $V$ as a function of the effective scalar field, $\phi$ driving inflation in the top and bottom panels respectively.}
\label{FIG1}
\end{figure*}

Of the $2\times10^4$ models generated, $568$ are consistent with a fiducial cosmological amplitude corresponding to $r = 0.05 \pm 0.001$ for modes that exited the horizon $60$ efoldings before the end of inflation.
Fig.~\ref{FIG1} displays the expansion rate, $H$, and the effective potential driving the expansion, $V$, for this subset in violet.  The left panels demonstrate the evolution of the expansion rate and potential, respectively, as they are plotted as a function of the time analog representing the number of efoldings before the end of inflation.  The right panels show the same quantities as a function of the effective scalar field driving inflation, thus illustrating the physical behavior of the expansion and energy during inflation.

The addition of direct detection constraints by an interferometer further decreases the allowed family of models.  A local scale measurement of  $\Omega_{\rm GW}(k_0 = 1.6\times10^{14}\,{\rm Mpc}^{-1}) = 8.2 \times10^{-17}$  corresponds to modes that were driven outside the causal horizon approximately $20$ efoldings before the end of inflation.  Models consistent with this fiducial amplitude within $8\%$ (the $2\sigma$ confidence interval for a DECIGO-like experiment with sensitivy scaled by $u=1/5$ at this fiducial amplitude) are displayed in Fig.~\ref{FIG1} in blue.  This additional local scale constraint reduces the number of allowed models to $93$, $16\%$ of those consistent with the future cosmological constraints and $0.5\%$ of those consistent with current measurements.

\begin{figure*}[t]
\includegraphics[width=6in]{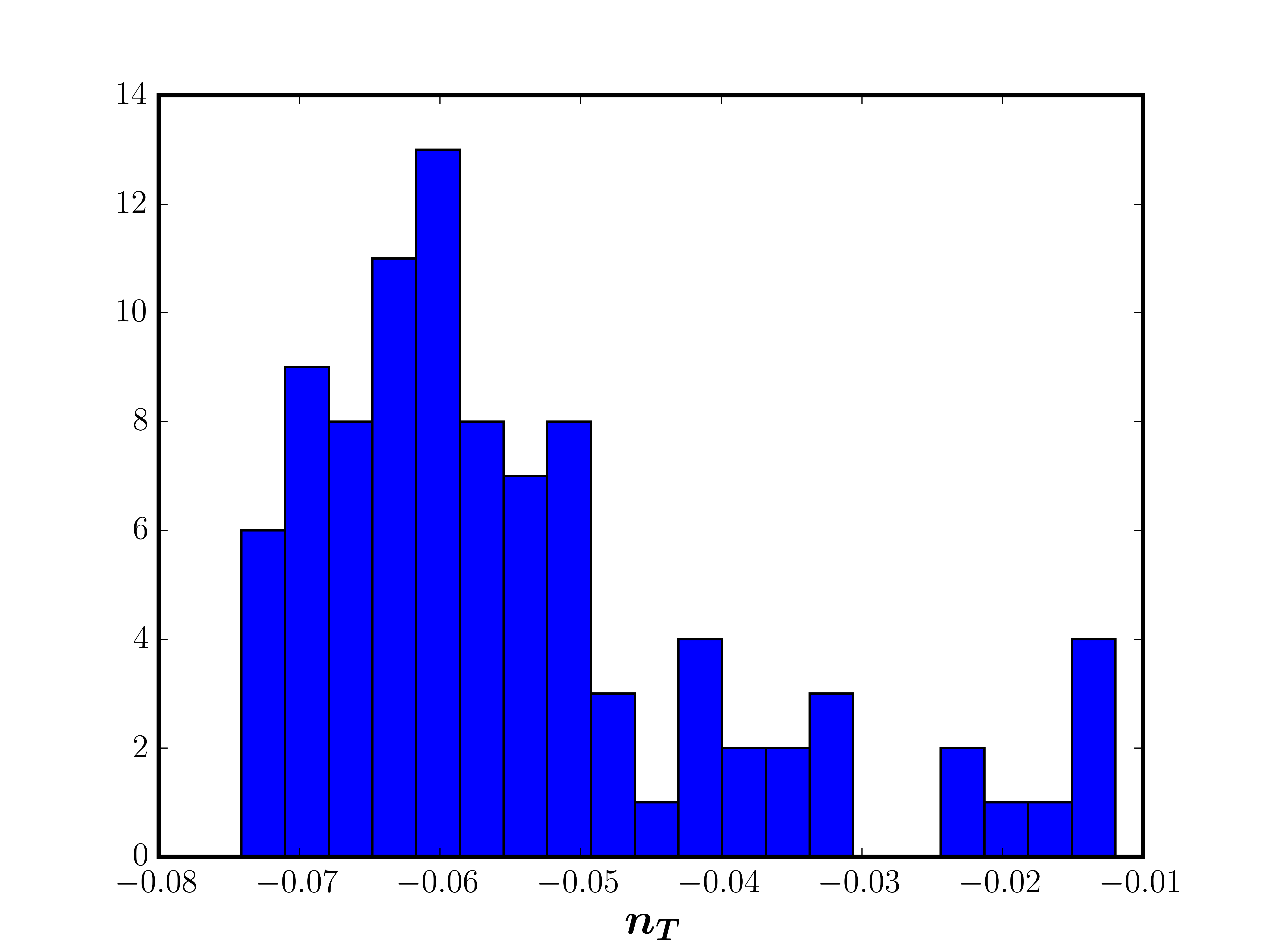}
\caption{The distribution of spectral tilts of models consistent with cosmological constraints with amplitudes corresponding to $r=0.05\pm0.001$ and at local scales with $\Omega_{\rm GW} = (8.2 \pm 0.69) \times10^{-17}$ at $95\%$ confidence for modes with wavenumber $k=1.6\times10^{14} Mpc^{-1}$ that exited the horizon $20$ efoldings before the end of inflation.}
\label{FIG2}
\end{figure*}

Qualitatively, all models consistent with constraints at cosmological and local scales show similar behavior, as illustrated by the blue curves in the left panels in  Fig.~\ref{FIG1} between $N=60$ and $N=20$.  A slight increase in the range of model behavior can be seen for $40 > N > 20$, which then translates to a widening of possibilities toward the end of inflation when the fields are departing from slow-roll conditions.  Without intermediate and late inflationary epoch constraints on the tensor spectral amplitudes, we can identify properties of the local scale spectrum in order to further constrain the family of models permissible by observation.  Specifically, the variety of allowable model amplitudes during intermediate and late time inflationary epochs will also reveal a distribution of spectral tilts, $n_T$, at local scales.  For example, the blue curves in Fig.~\ref{FIG1}, demonstrate a distribution of spectral tilts, $n_{\rm T}$, for modes with wave numbers of $k=1.6\times10^{14} \rm Mpc^{-1}$ that exited the horizon $20$ efoldings before inflation ended.  This distribution is shown in Fig.~\ref{FIG2}.

A group of allowed models are clustered around spectral tilts of $n_T=-0.06$ 
at local scales. The tail in the distribution toward higher values of $n_{\rm T}$ is easily associated with the models in Fig.~\ref{FIG1} that fall more rapidly during intermediate epochs and then require a plateauing behavior in order to satisfy the required number of efoldings of expansion.  These models all lie below the simpler main group of curves during intermediate times and above toward the end of inflation.  These outlying models can be ruled out if interferometric measurements obtained sensitivity levels to constrain the tilt to $n_{\rm T}= - 0.06 \pm 0.02$ or better.  In order to achieve these constraints at $95\%$ confidence, the previously discussed DECIGO experiment will need a sensitivity level scaling parameter $u = 1/25$, an additional factor of $5$ below the strain sensitivity (or a factor of $25$ below the sensitivity to $\Omega_{\rm GW}$) used above ($u = 1/5$). This is a factor of $625$ increase in sensitivity to $\Omega_{\rm GW}$ beyond the base design for DECIGO.

\begin{figure*}[t]
\includegraphics[width=6in]{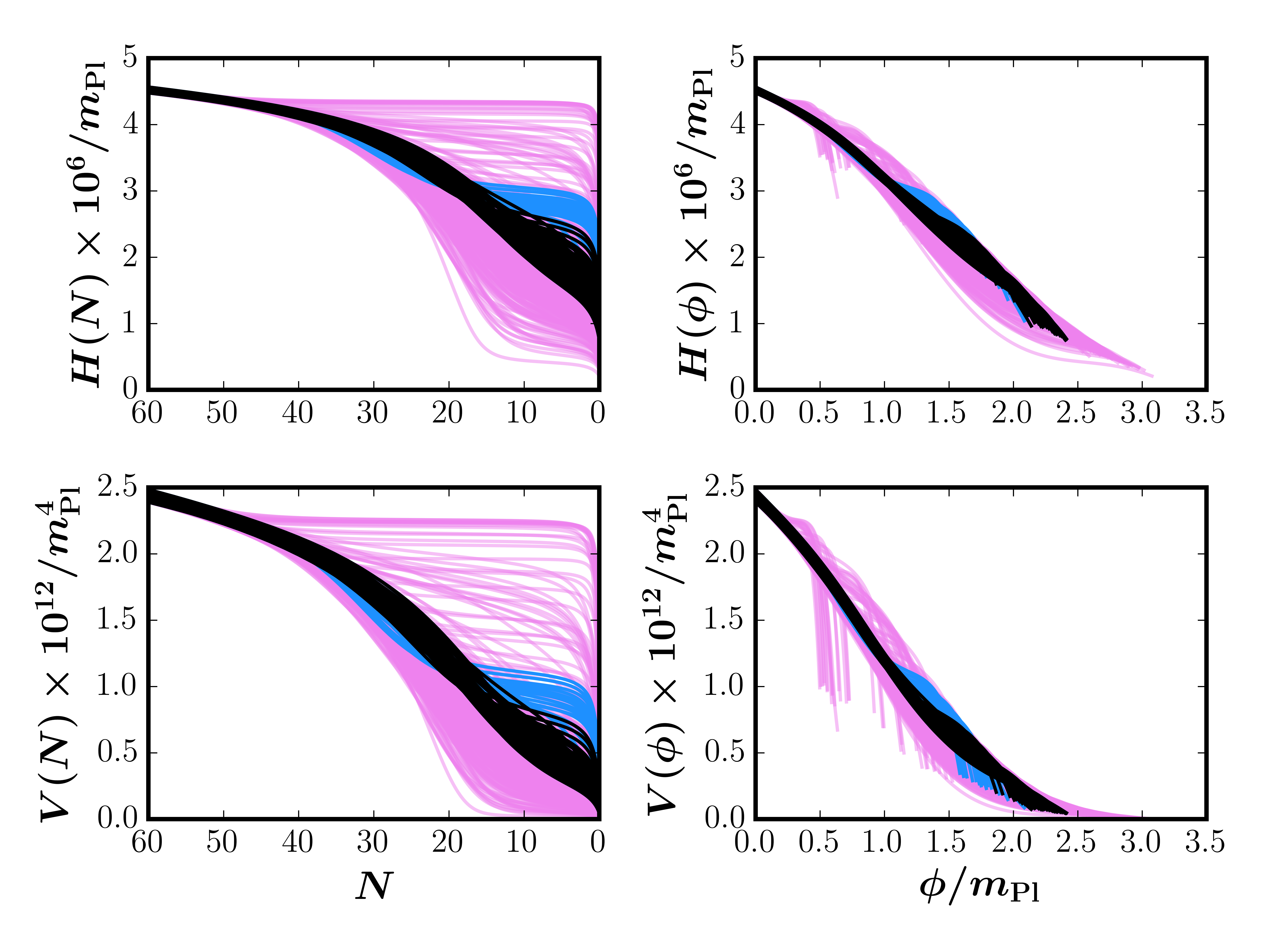}
\caption{The same models as shown in Fig.~\ref{FIG1}. The smaller subset of models that have local scale tensor spectral tilts of $n_{\rm T} \in [-0.08,-0.04]$ are plotted in black.}
\label{FIG3}
\end{figure*}

\begin{figure*}[t]
\includegraphics[width=6in]{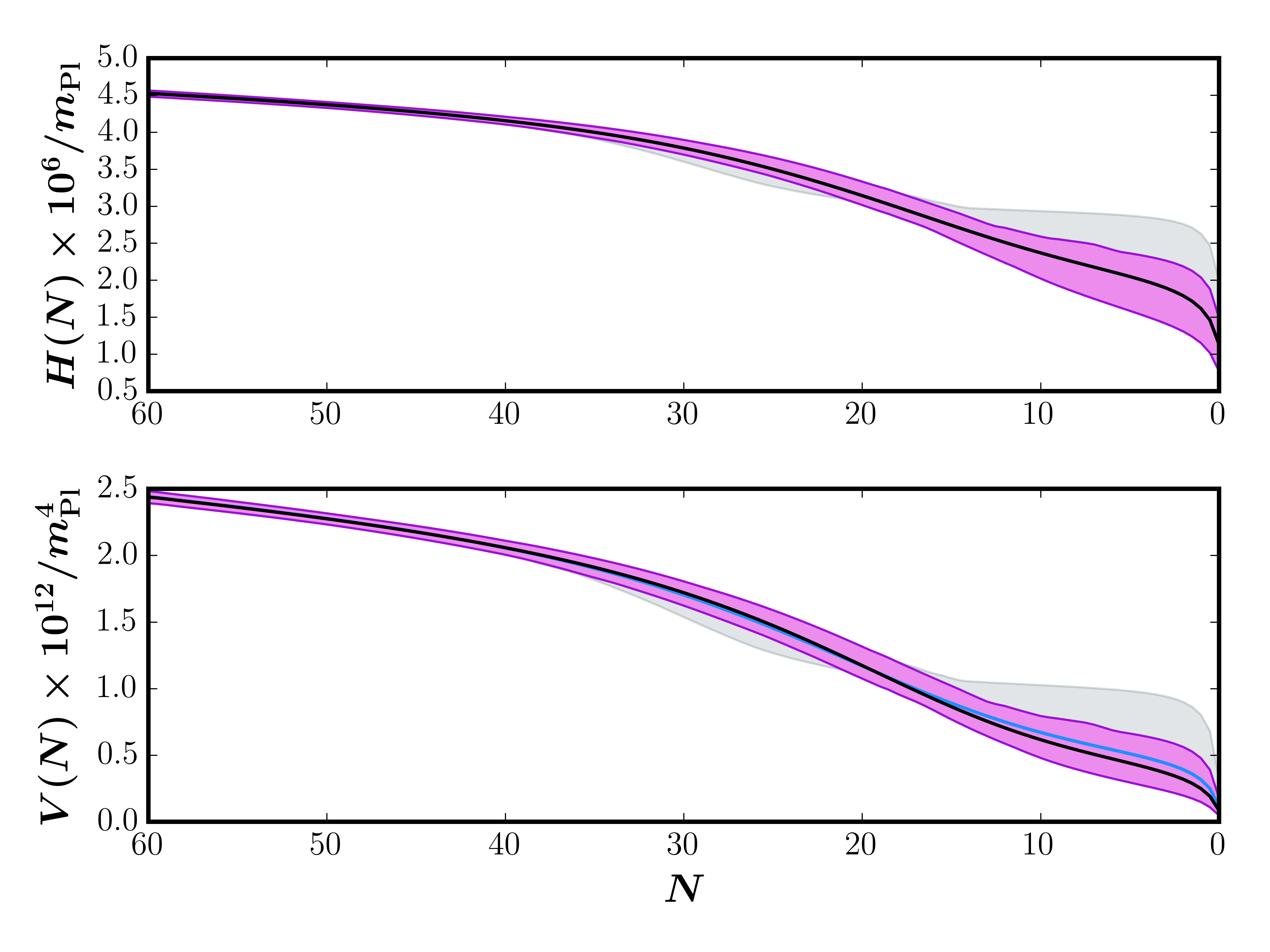}
\caption{The mean and $95\%$ spread about the mean of models shown in Fig.~\ref{FIG3}. Here the potentials and expansion rates consistent with $r=0.05\pm0.001$ with $2\sigma$ confidence at cosmological scales (corresponding to $60$ efoldings before the end of inflation), as well as at local scales (corresponding to 20 efoldings before the end of inflation) with $8\%, 2\sigma$ error on the amplitude have their mean plotted in blue and the spread of models illustrated as a gray band.  The set of models that additionally have spectral tilt constraints at local scales in the range of $-0.08 < n_{\rm T} < -0.04$ have their mean plotted in black with their spread about the mean as a violet band.}
\label{FIG4}
\end{figure*}

Applying these spectral tilt constraints to the models consistent at both cosmological and local scales as shown in blue in Fig.~\ref{FIG1} (albeit without constricting the constraints on local scale amplitudes), we further reduce the number of consistent models to 78.  This subset is illustrated in Fig.~\ref{FIG3} in black, plotted over the models shown in Fig.~\ref{FIG1}.  To further illustrate the distribution of each of these subsets of models, Fig.~\ref{FIG4} indicates the mean and the band containing $95\%$ of the models about the mean for each family of curves shown in Fig.~\ref{FIG3}.  Here, the blue curves from Fig.~\ref{FIG3} have their mean plotted as blue and the $95\%$ spread of models shown as gray.  Similarly for the violet curves from Fig.~\ref{FIG3}, the mean and $95\%$ spread of models about the mean are shown in Fig.~\ref{FIG4} as black and violet respectively.  Although this additional constraint has not greatly reduced the number of allowed models,it has significantly restricted the qualitative form of the permissible models. If the local scale spectral index were measured to be $n_T=-0.02\pm 0.02$, only a few models with more complex potential shapes would be consistent.

This conclusion will hold even for a tensor spectrum with a lower amplitude than the fiducially chosen value, provided that the amplitude can still be measured on both scales with similar fractional errors as used here. Of course, the lower the amplitude, the more technically challenging these measurements become. For amplitudes significantly below $r=10^{-3}$, confusion with the residual gravitational lensing signal will prevent measurement of the tensor amplitude using B-mode polarization \cite{Knox:2002pe,Hirata:2003ka}, while the binary confusion limit becomes an increasing problem for interferometer measurements.
 
\section{IV. Discussion}
These results clearly demonstrate the capacity of combined measurements of the tensor power spectrum at both microwave background and interferometric scales to constrain the history and physics of inflation. The family of inflation models consistent with both measurements is vastly restricted compared to inflation models consistent only with an amplitude measurement at a single scale. Furthermore, the combined measurements determine the inflaton potential and expansion history over a wide
range of field values and evolutionary epochs. We have assumed B-mode polarization measurements of the tensor amplitude with precision at the $r=0.001$ level; upcoming polarization experiments with increased sensitivity and frequency coverage are expected to surpass this (see, e.g., \cite{Kogut:2011xw}).  We also assume a future interferometer measurement of the tensor amplitude at the $8\%$ level, which will be challenging but feasible if there is in fact a measureable cosmological tensor signal.  Pushing the sensitivity of interferometric experiments farther opens the possibilty of further restricting the family of allowed models via the tilt of the spectra at these scales.  Clearly such a measurement vastly restricts the behavioral variety of consistent inflationary models. 

It may seem surprising that the tilt of the tensor spectrum varies so greatly between scales corresponding to $60$ and $20$ efoldings before the end of inflation. Recall, however, that during slow-roll $\epsilon \propto (V'/V)^2$, where the prime indicates the derivative with respect to the effective scalar field $\phi$ and $V$ is the effective potential of that scalar field. As illustrated in Fig.~\ref{FIG1} and \ref{FIG3}, the potentials drop by a factor of approximately $3$ between $N=60$ and $N=20$. When plotted as a function of $\phi$, this roughly corresponds to the interval of $\phi/m_{\rm Pl} \in [0.0,1.0]$.  In this interval, the slope of the potential steepens only slightly. Taking both the slight steepening of the slope and the drop in the value of the potential into consideration, we estimate epsilon to increase by a factor of \emph{at least} $9$. Our fiducial selection of models at $N=60$ have amplitudes corresponding to $r=0.05\pm0.001$ and have tilts centered on $n_{\rm T} = -6.25\times10^{-3}$ (as to be expected according to the slow-roll consistency relation, $r=-8n_{\rm T}$).  Moving to $N=20$ and increasing $\epsilon$ by a factor of roughly $9$ then results in a distribution of tilts at these scales to be centered on roughly $n_{\rm T} \approx -0.06$. This is consistent with what we have found in the distribution of tilts at local scales, as illustrated in Fig.~\ref{FIG2}, and is indicative of the dominance of slow-roll behavior throughout these $40$ efoldings of evolution.  The outlying models underwent brief departures from slow-roll evolution before returning to ensure the necessary number of efoldings of expansion are completed. This behavior directly translates to the tilt of the spectrum at local scales, thus providing a probe of the expansion history and effective potential driving inflation during epochs corresponding to otherwise unmeasureable scales.

We have made the simplifying physical assumption that reheating after inflation happens quickly, i.e. on a time scale less than a Hubble time, so it has little effect on the expansion history of the Universe. Some models of reheating take significantly longer than this, resulting in a period of matter-dominated expansion prior to the usual radiation-dominated era which can modify the tensor amplitude at small scales (e.g., \cite{Kuroyanagi:2010mm}). Measurements of both the tensor amplitude and the scalar spectral index will give interesting constraints on the duration of any reheating epoch \cite{Dai:2014jja}. An extension of the present analysis to include reheating will be considered elsewhere. 

The numerical analysis in this paper does not have a rigorous quantification of Monte Carlo coverage of the inflation model space. The truncation of the slow-roll hierarchy at a given order results in only a particular subset of inflation models in the allowed model space. Larger computational efforts can include higher-order truncations, effectively expanding the model space which is explored, and more total models in each Monte Carlo, which will sharpen statistical conclusions. It is unlikely that models not available in our 8th-order slow-roll hierarchy will change our conclusions in a qualitative sense, since the allowed models span a wide range of potentials, as is visible in Fig.~\ref{FIG1}. However, whether rare successful models exist which are clearly different from the family of models identified here is an interesting open question. These rare cases may be further constrained by additional sources of information.

The results presented here show that a detection of inflationary tensor perturbations at two widely separated scales, the cosmological scale via B-polarization of the microwave background and the Earth-Moon scale via a space-based laser interferometer, will determine the dynamical history of the Universe during the inflation era. In turn, such measurements will give strong constraints on fundamental physics at energy scales of $10^{16}$ GeV, inaccessible by any other means.

\acknowledgments
A.K.\ thanks B.~Fishbain for helpful suggestions. J.C. thanks J. Newman for access to computational power as well as S. Aiola and A. Congdon for helpful conversations.
J.C.\ and A.K.\ are supported in part by the National Science Foundation
under grant NSF-AST-1312380. 
W.H.K.\ is supported by the National Science Foundation under grants NSF-PHY-1066278 and 
NSF-PHY-1417317.  N.S.\ is supported by MEXT grant 24103006.
Computations were performed in part at the University at Buffalo Center for Computational Research.
Flowcode 1.0.3, used to perform the
inflation flow Monte Carlo, was written by Brian A. Powell and W.H.K. 
This work has used NASA's Astrophysical Data System for bibliographic information.

\bibliography{constraints_prl_submit}


\end{document}